\documentclass[epj]{svjour}
\usepackage{amsfonts,amssymb,amsmath}
\usepackage{latexsym}
\usepackage{cite}
\usepackage{bm}
\usepackage{graphicx,afterpage}

\begin{document}
\title{Vorticity Banding During the Lamellar-to-Onion Transition in a
  Lyotropic Surfactant Solution in Shear Flow}
\author{Georgina M. H. Wilkins  \and Peter D. Olmsted}
\institute{School of Physics and Astronomy, University of Leeds,
Leeds LS2 9JT, United Kingdom }
\date{Received: \today / Revised version: date}
\abstract{We report on the rheology of a lamellar lyotropic surfactant
  solution (SDS/dodecane/pentanol/water), and identify a discontinuous
  transition between two shear thinning regimes which correspond to
  the low stress lamellar phase and the more viscous shear induced
  multi-lamellar vesicle, or ``onion'' phase. We study in detail the
  flow curve, stress as a function of shear rate, during the
  transition region, and present evidence that the region consists of
  a shear banded phase where the material has macroscopically
  separated into bands of lamellae and onions stacked in the vorticity
  direction.  We infer very slow and irregular transformations from
  lamellae to onions as the stress is increased through the two phase
  region, and identify distinct events consistent with the nucleation
  of small fractions of onions that coexist with sheared lamellae.
\PACS{
      {83.}{Rheology}   \and
      {83.60.Wc}{Flow instabilities} \and
      {83.80.Qr}{Surfactant and micellar systems, associated polymers}
     } 
} 

\authorrunning{Wilkins and Olmsted}
\titlerunning{Vorticity Banding in Lamellar Phases under Shear}
\maketitle

\section{Introduction}\label{sec:0}
Surfactant lamellar phases have a wide range of flow - induced
behaviour including shear thickening
\cite{RND93,Bergenholtz_Langmuir_1996}, shear thinning
\cite{RND93,Diat_JPhys_II_93_2}, shear induced structures
\cite{RND93,Bergenholtz_Langmuir_1996,DRN93b} and shear-induced
phase separation
\cite{RND93,Bergenholtz_Langmuir_1996,Hoff94,DRN93b,FMG95,Pani+98,Kleman96,zipfel}.
In one of the most dramatic transitions, discovered by Diat and
Roux, shear flow can transform some lamellar phases can into
multi-lamellar vesicles, or ``onions"
\cite{RND93,Diat_JPhys_II_93_2}.  In the SDS/ dodecane/ pentanol/
water system, stronger shear flow eventually destroys the onion
phase in favour of either a highly aligned lamellar phase or
multicylindrical ``leek-like'' structure
\cite{wunenberger_EurPhysJE_00}. This is an example of a shear
induced microstructural phase transition. Other microstructural
transitions have also been reported for lamellar phases made from
different ingredients. For example, an SDS/ octanol/ brine mixture
exhibits a transition from the lamellar to the sponge phase at
$30^{\circ}$C\cite{Herve93}. At low imposed shear stresses the same
system features a transition from a liquid onion phase to an ordered
phase of onions organised into flat planes sliding over one another
\cite{Sierro95}.  Conversely, for high enough shear rates, a certain
SDS/ decanol/ water mixture undergoes a transition as a function of
increasing decanol fraction, from the $\hat{c}$-orientation,
consisting of lamellae that slide over each other with layer normals
parallel to the flow gradient direction, to the
${\hat{a}}$-orientation, consisting of lamellae that lie in the
shear plane with layer normals in the vorticity direction
\cite{Zipfel99}. Interesting kinetics have been seen in other
lyotropic lamellar phases, such as cylindrical or leek-like
intermediates during the transition from lamellae to onions
\cite{zipfel}. In this work we revisit the transition originally
found by Diat and Roux in SDS/dodecane/pentanol/water, and study the
lamellar-to-onion (L-O) transition region in more detail.

The onion phase is of great interest to the chemical industry:
onions form in a variety of different surfactants and lipids, and
the size is easily controlled by the magnitude of the applied shear
rate $\dot{\gamma}$ or shear stress $\sigma$.  As a result, onions
can be designed for specific tasks such as micro-encapsulation of
drugs or colour pigment.  From an academic viewpoint, the
shear-induced transition from lamellae to onions is a
topology-changing transition that is still poorly understood,
despite growing knowledge of defects in lamellar and condensed media
\cite{klemanbook}.  Unresolved issues about the L-O transition
include the mechanism of instability, physical nature of the
transformation, possible phase coexistence between sheared lamellae
and onions, the compatibility between onions and lamellar
orientation, and the role of dislocations.  We will argue below that
the L-O transition is an example of ``shear banding'', or separation
of material into macroscopic bands of material. In this case the
signatures are consistent with bands stacked along the vorticity
direction, rather than the more conventionally seen layering in the
flow gradient direction \cite{olmsted99c}.

\begin{figure*}[htb]
  \begin{center}
    \includegraphics[scale=0.5]{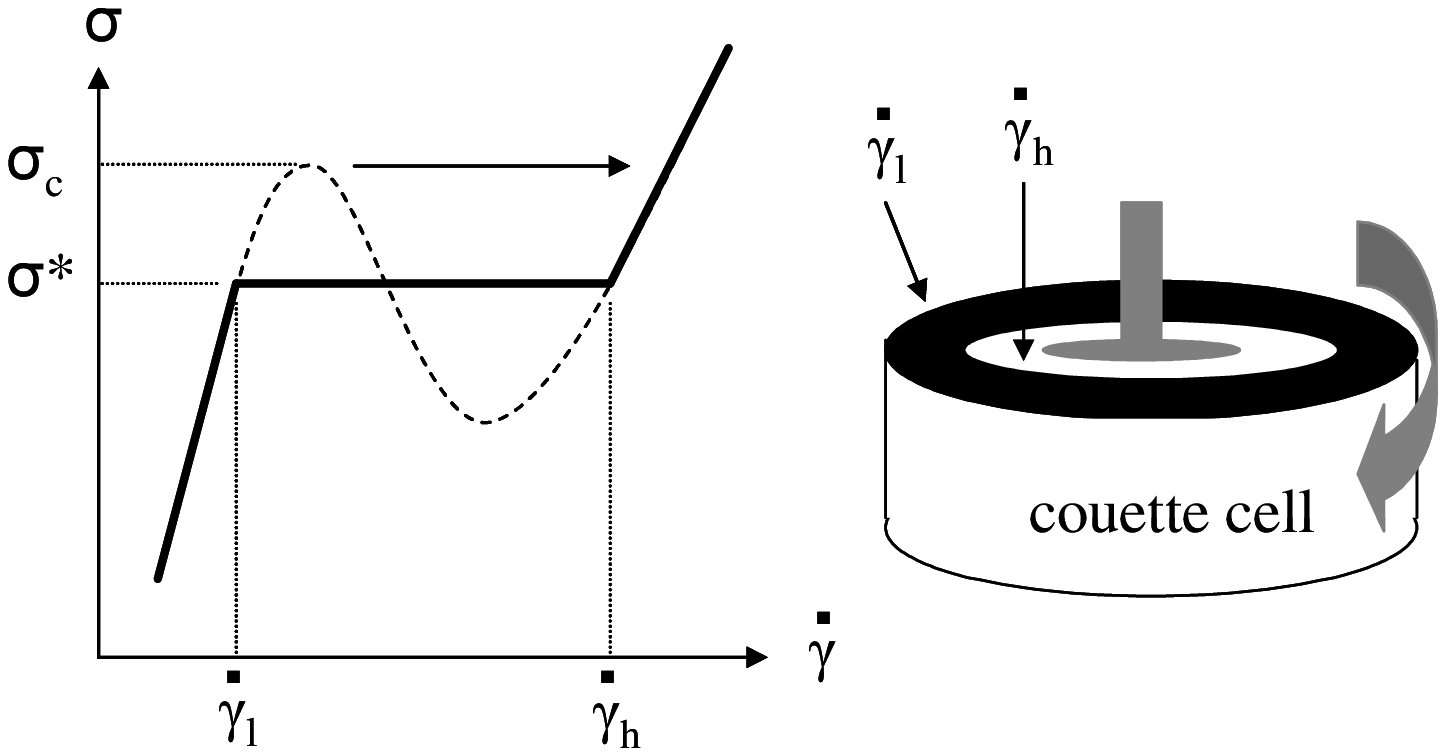}
    \includegraphics[scale=0.5]{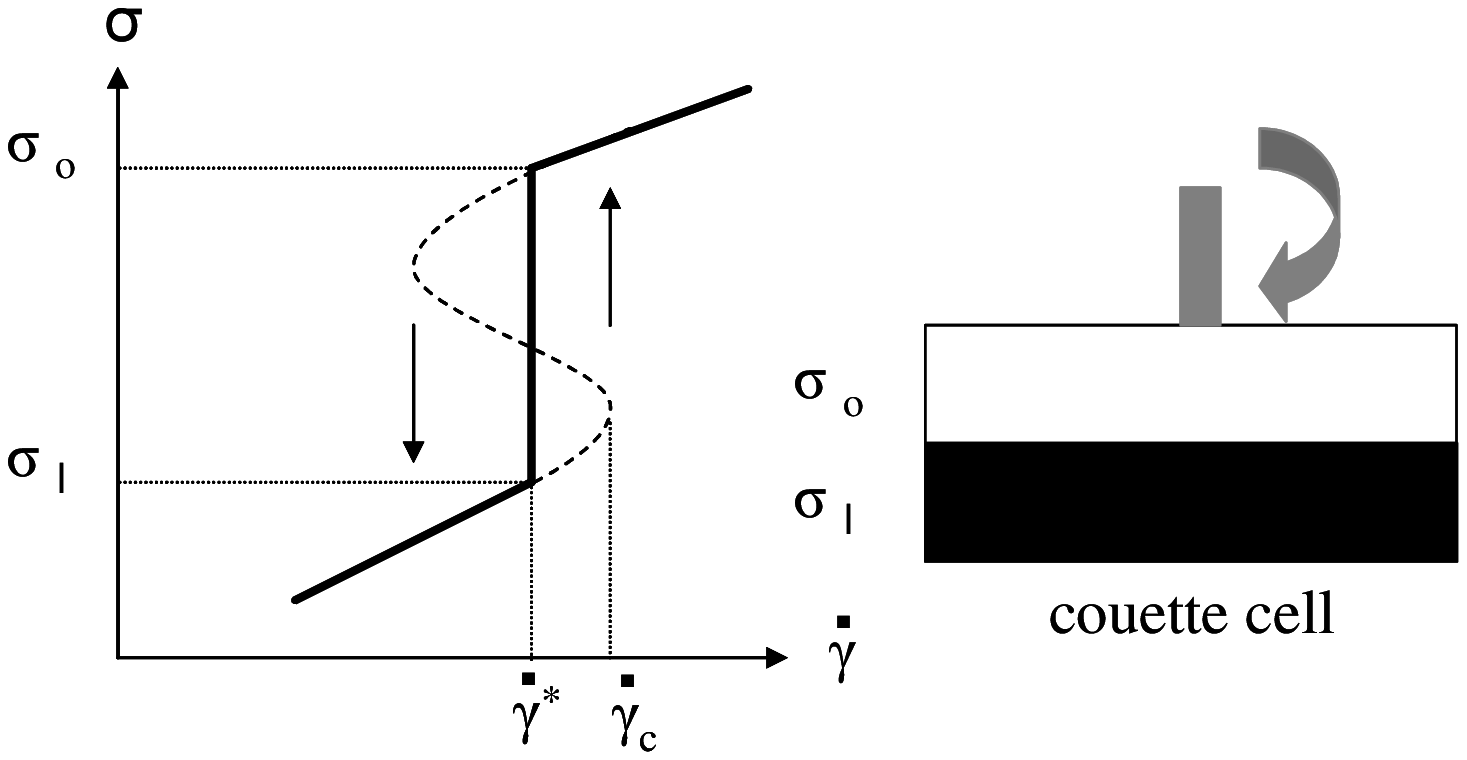}
    \caption{Schematic flow curves for shear banding in cylindrical
      Couette geometry. Left: gradient banding, in which a high shear
      rate band flows near the inner cylinder and a low shear rate
      band flows near the outer cylinder.  In ideal gradient banding
      the two phases coexist under imposed strain rate conditions,
      along the plateau at a selected shear stress $\sigma^{\ast}$.
      Right: Vorticity banding, in which the shear bands are stacked
      in the vorticity direction; in the ideal scenario shear bands
      coexist under controlled stress conditions, on a cliff at a
      selected strain rate $\dot{\gamma}^{\ast}$ \cite{Olms99}.  In
      both cases the solid line shows the composite flow curve that
      would be measured in ideal shear banding, while the dashed line
      shows metastable (positive slope) or unstable (negative slope)
      portions of the constitutive curve. The arrows indicate some
      possible metastable pathways that could be followed for rapid
      experiments: such as rapidly increased stress in gradient
      banding, or rapidly decreased or increased strain rate in
      vorticity banding.}
    \label{fig:bandcartoon}
  \end{center}
\end{figure*}

The best studied examples of shear banding are in solutions of
entangled wormlike micelles in strong shear flows
\cite{Khat+93,rehage91,Cate90,Cate96}.  Flow induces an instability to
a well-aligned and high viscosity state, such that under imposed
strain rate conditions the fluid ``phase separates'' into macroscopic
regions of material flowing at different shear rates, along a stress
plateau that spans a range of shear rates
(Fig.~\ref{fig:bandcartoon})\cite{SCM93}. The shear-induced band of
lower viscosity material increases in size as the average shear rate
is swept across the plateau, and the shear bands are separated along
the flow gradient direction (\textit{gradient shear banding}),
consistent with a common stress between bands. Gradient shear banding
has been observed in dilute wormlike micelles using techniques such
as magnetic resonance \cite{Call+96}, ultrasonic velocity profiling
\cite{Becu04} and optical birefringence microscopy \cite{lerouge}.

Shear bands can also form along the vorticity direction
(Fig.~\ref{fig:bandcartoon}) \cite{OlmsLu97}.  This condition could
be realized when two states, such as a fluid and induced gel, have
multiple stresses for a range of strain rates. Phases with
coexisting shear stresses do not violate the momentum balance as
long as neighbouring bands are separated in the vorticity direction,
although a normal stress condition must still be satisfied to
determine the coexisting strain rate $\dot{\gamma}^{\ast}$. Thus,
vorticity shear bands have different shear stresses whose sum must
equal the total applied stress.  For a shear thickening transition
the corresponding rheological flow curve of such a material would
exhibit a characteristic step increase of shear stress at a critical
shear rate \cite{Olms99}, as shown in Fig.~\ref{fig:bandcartoon}.

The original work of Diat and Roux \cite{Diat_JPhys_II_93_2} showed
a vertical step at a critical shear rate (in their most concentrated
surfactant bilayer solution, $\phi_{oil}=0.50$), which suggests that
the SDS system is a candidate for vorticity banding. They also
reported seeing structures along the vorticity direction, but in the
range of shear rates associated with the transition between onions
and the higher shear rate phase (either multicylindrical ``leeks''
or well aligned lamellae). Vorticity banding associated with a
thinning transition, the alignment of a polydomain colloidal crystal
in solution, was also been observed but not noted as such by Chen
and Zukoski \cite{ChenZAHSBG92,ChenCAZ94}.  More recently, Callaghan
and co-workers, using magnetic resonance imaging, observed what is
apparently a combination of vorticity and gradient banding in a
semidilute wormlike micelle solution in a cone and plate geometry
\cite{callaghan}. Fisher and co-workers have also observed structure
formation along the vorticity direction in another wormlike micelle
solution \cite{WFF98}.

Here we study the shear induced microstructural L-O transition in the
SDS system first studied by Diat and Roux, focusing on vorticity shear
banding signatures. The paper is organised as follows. In section 2 we
briefly describe the system and the experimental setup for measuring
the rheology. In section 3, we summarise the homogeneous flow
behaviour of the lamellar phase beginning with the low stress yield
behaviour. In section 4 we study the macroscopic nature of the
lamellar to onion transition and investigate effects of experimental
protocol on the existence of the transition region. We finish with a
discussion.

\section{System and Experimental Details}\label{sec:1}

\subsection{Sample preparation}

The lyotropic lamellar phase under study is a quaternary mixture of
sodium dodecyl sulphate (SDS), pentanol, dodecane and water. The
materials were supplied by Aldrich Chemicals and used without
further purification. Millipore multi-Q water with a resistivity
better than 18.0 MOhm cm was used throughout.  The lamellar phase is
most easily obtained by diluting a concentrated lamellar phase
(weight fractions of 47.4\% water, 22.0\% pentanol and 30.6\% SDS)
with an oil mixture (weight fractions of 92.0\% dodecane and 8.00\%
pentanol). The lamellar phase exists along a very wide range of this
dilution line. In Ref.~\cite{Diat_JPhys_II_93_2} these two mixtures
were first prepared separately and then mixed. Since this method
often resulted in an inhomogeneous solution we mixed the required
proportions of the constituent materials directly by shaking
vigorously for several minutes. After shaking, the solution was
placed in an oven at $60.0^{\circ}\textrm{C}$ for several days.

At equilibrium, this phase comprises layers of water surrounded by
surfactant separated by dodecane.  Pentanol acts as a co-surfactant
and occupies the oil part of the phase. At room temperature, the
solution forms a lamellar phase for a large range of dodecane
concentrations ($0.466<\phi_{oil}<0.644$, where $\phi_{oil}$ denotes
the weight fraction of dodecane in the mixture) \cite{safinya}. The
bilayers are stabilised by thermal undulations \cite{rouxsafinya88},
with a layer spacing $d$ ranging from 5 nm to 20 nm depending on the
concentration of the phase\cite{Diat_JPhys_II_93_2}. In this work we
investigate lamellar phases with dodecane concentrations
$\phi_{oil}=0.466, 0.50, 0.569, 0.644$.
\subsection{Rheology}
A Rheometrics Scientific SR500 stress controlled rheometer was used
throughout. A cone and plate geometry with a 40 mm diameter cone with
angle $\alpha = 0.02$ radians was mounted on the rheometer stress
head. A home-made Perspex solvent trap was used to minimize solvent
evaporation. The plate was thermostatted using a water cooling system,
and all experiments were performed at $24^{\circ}$C.

Sweep experiments are performed by imposing an initial stress after
sample loading. The shear stress range is $0.1\,\mathrm{Pa} < \sigma
< 100\,\mathrm{Pa} $. For larger stresses, depending on the
concentration, the sample often spurted out of the cell, with the
more viscous lamellar phases (lower $\phi_{oil}$) being ejected at a
lower stress. During the sweep experiments, successive stresses were
imposed for either (a) a prescribed time $\tau_{max}$ or (b) until
steady state criteria are reached. These criteria (or steady state
test) identify an early steady state where the steady state is
defined as the point where the shear rate $\dot\gamma$ remains
within an ``acceptance window''.

In the steady state test the measured shear rate is monitored as a
function of the elapsed time since the previous stress increment. The
steady state is defined according to the RMS fluctuations $P$ of the
shear rate around the mean shear rate,
\begin{equation}
  P = \frac{\sqrt{\langle\left(\dot{\gamma} -
    \bar{\dot{\gamma}}(t)\right)^2 \rangle_{5\%}}}{\bar{\dot\gamma}}
\end{equation}
where the angle brackets and overbar denote a time average over the
previous 5\%($0.05t$) of the current duration $t$ of the test.
Typically the condition $P=1\%$ or $P=2\%$ was used. If the sample
does not fall below set value for $P$ within the maximum delay time
$\tau_{max}$, then the stress is incremented to the next stress in the
series.  Unless otherwise stated, $\tau_{max}=2000\,\textrm{s}$.

\section{Non-linear rheology}\label{sec:2}
Previous work on this system suggests the following succession of
phases as a function of shear rate for the SDS/do\-decane/pen\-ta\-nol
mixture
\cite{RND93,Diat_JPhys_II_93_2,Pani+98,Cristobal.Rouch.ea01,Courbin.Delville.ea02,Courbin_PhysRevE_04}.
An aligned smectic with no defects is solid like for flows that deform
the layer spacing or induce bending, and fluid for flows that induce
only layer sliding or in-layer flows. In principle a highly defected
phase is solid-like with a yield stress due to defects; the yield
stress is expected to be history dependent
\cite{Colby_EuroPhysLet_01}. Upon applying a shear stress above the
yield stress the layered phase flows, with layer alignment generally
in the $\hat{c}$ orientation, in which layer normals are parallel to
the flow gradient direction \cite{Diat_JPhys_II_93_2}. The system
flows until a critical shear rate or shear stress is reached (see
below for a discussion), at which point the flowing lamellar phase is
thought to undergo an undulation instability similar to the
Helfrich-Hurault instability
\cite{RND93,marlow02,wunenberger01,zilman99,Courbin.Delville.ea02} to
an onion phase. Candidate mechanisms for this undulation instability
include local dilational stresses due to defects or wall asperities
\cite{RND93}, ``wrinkling out'' of fluctuations into long wavelength
modes in materials with no defects \cite{zilman99}, or a change in
preferred layer spacing due to reduced collisions, which leads to an
effective dilational strain \cite{marlow02}. Following a discontinuous
step in the shear rate, the sample is entirely converted to onions.

When the sample reaches a steady state the onions are monodisperse
with radii $R$ of the order of a few microns \cite{RND93,meyer}. The
size can be heuristically understood as a balance of elastic
deformation and viscous forces between onions \cite{RND93},
\begin{equation}\label{R_dependence}
R=\sqrt{\frac{4{\pi}(2\kappa+\bar{\kappa})}{d\eta \dot{\gamma}}}
\end{equation}
where $\kappa$ and $\bar\kappa$ are the mean and Gaussian curvature
moduli of the layers, $\eta$ is a viscosity, and $d$ is the smectic
layer spacing.  This approach works best when the viscosity used is
the solvent viscosity rather than the apparent viscosity
(\textit{i.e.} flow only occurs between
onions)\cite{Courbin_PhysRevE_04}.  Note that this argument does not
address the mechanism by which onion sizes adjust, which involves
other processes such as peeling and accretion of layers, internal
layer collapse in the onions due to high internal stresses, and
expulsion and absorption of solvent\cite{Leon_JCondMat_02}. Panizza
and co-workers measured the size to be
\begin{equation}
  \label{eq:R}
  R=R_0\left(\frac{\dot{\gamma}_0}{\dot{\gamma}}\right)^{-1/2},
\end{equation}
where $R_0=5.8\,\mu\mathrm{m}$ and
$\dot{\gamma}_0=1\,\mathrm{s}^{-1}$ when the dodecane oil
concentration was $\phi_{oil}=0.45$ \cite{Pani+98}. The onions
decrease in size with increasing shear rate, and at higher shear
rates are unstable with respect to a well-aligned lamellar phase
with few defects. The flow curves are consistent with a gradient
banding coexistence of states along a stress plateau, although this
coexistence has not been well-studied and the flow curves were
measured using imposed stresses resulting in a discontinuity between
the onion flow branch and the well aligned lamellar flow branch
\cite{Diat_JPhys_II_93_2}.

\begin{figure*}[htb]
  \begin{center}
    \includegraphics[scale=0.55]{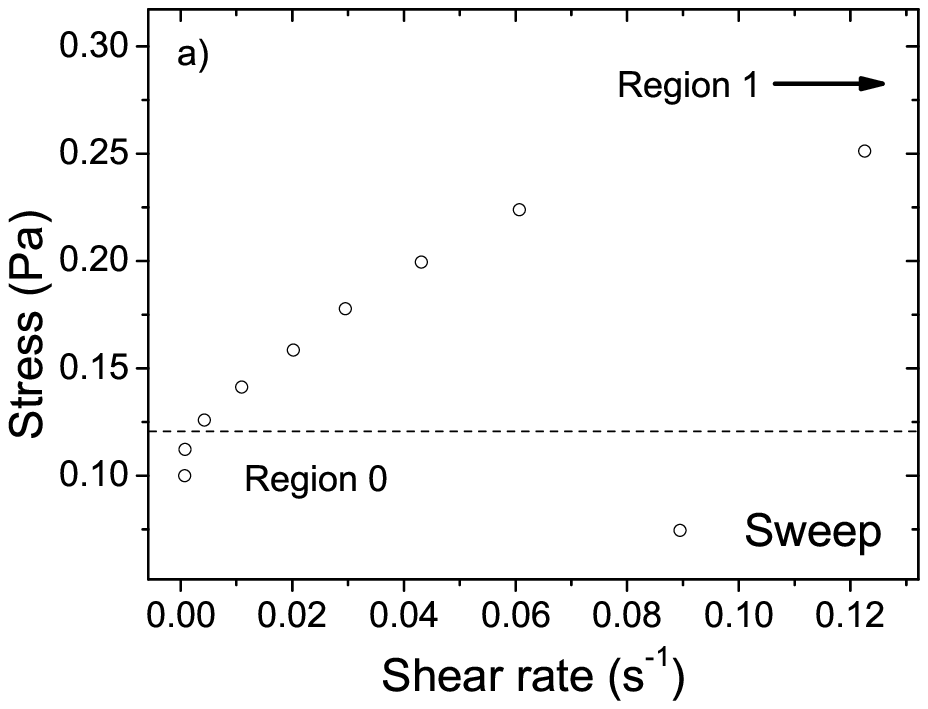}
    \includegraphics[scale=0.55]{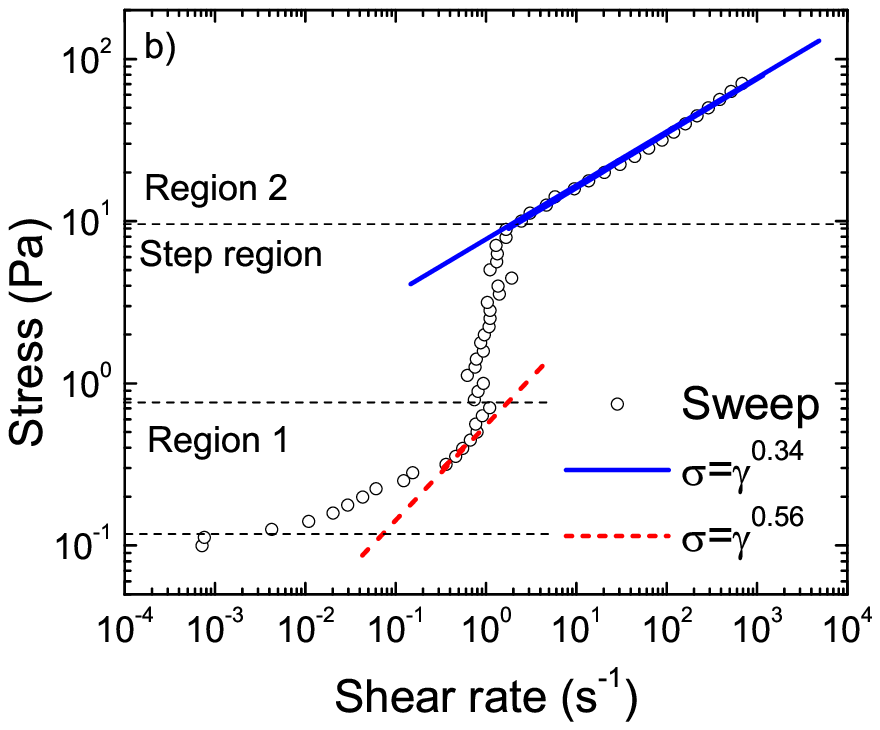}
    \includegraphics[scale=0.55]{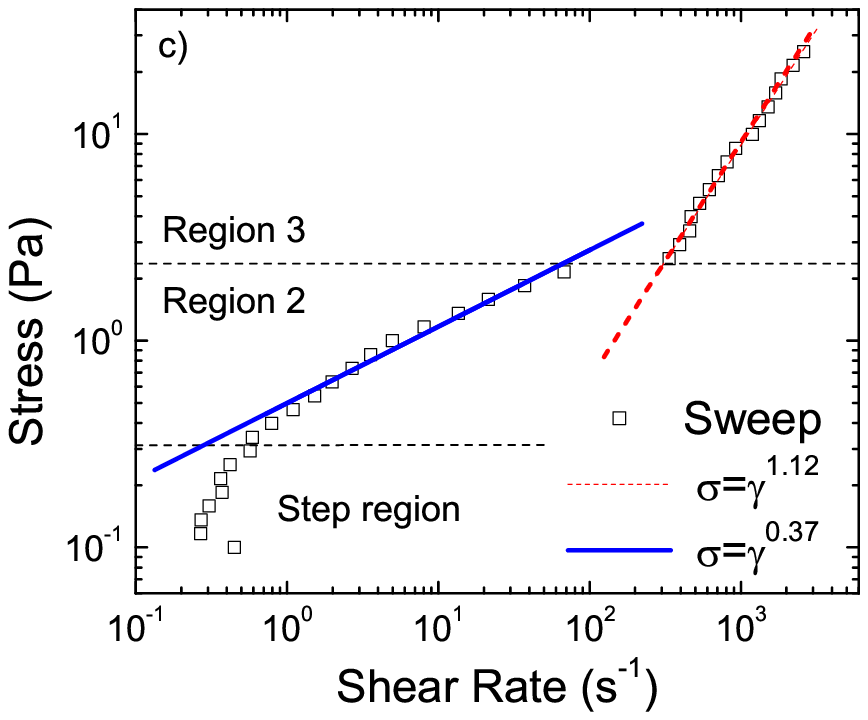}
    \caption{Stress sweeps using the steady state criteria
      $\tau_{max}=2000\,\textrm{s}$ and $P=1\%$. Power laws are used
      to fit the different regions: a) and b) Stress sweep for
      $\phi_{oil}=0.466$ ($\circ$).  The material shows little
      deformation at imposed stresses below the yield stress
      ($\sigma_y=0.12$ Pa).  c) Stress sweep for the more dilute
      lamellar phase $\phi_{oil}=0.644$ ($\scriptstyle \square$).}
      \label{Fig2_FlowRegions}
\end{center}
\end{figure*}
To illustrate these different regimes, figure~\ref{Fig2_FlowRegions}
shows the flow curves measured using the sweep experiment protocol
described above for oil concentrations $\phi_{oil}=0.466$
(Fig.~\ref{Fig2_FlowRegions}a and \ref{Fig2_FlowRegions}b) and
$\phi_{oil}=0.644$ (Fig.~\ref{Fig2_FlowRegions}c). The apparent
yield stress for the most concentrated membrane phase
($\phi_{oil}=0.466$, which corresponds to a layer spacing
$d\simeq8.2\,\textrm{nm}$ \cite{Courbin.Delville.ea02}) is
$\sigma_Y\simeq0.13$\,Pa. Meyer \textit{et al.} argued that the
yield stress is that stress needed to deform a network of
dislocations with average spacing $\xi$ \cite{meyer}:
\begin{equation}
  \label{eq:yield}
  \sigma_Y\sim \frac12\frac{\bar{B} \bar b^3}{\xi^3},
\end{equation}
where $\bar{B}$ is the smectic compression modulus, and $\bar b\simeq
b/(2\pi)$, where $b\simeq d$ is the Burger's vector of the
dislocation. This leads to an estimated defect spacing of $\xi=86\,$nm
with $\xi/d\simeq11$ membranes between defects.  Meyer and co-workers
estimated $\xi /\bar{b}=65$ for a thermotropic system (8CB), and $\xi
/\bar{b}=8$ for a lyotropic system (quasi-ternary ammonium surfactant
cetylpyridinium chloride, hexanol and brine), quite close to the value
we find in the SDS solution.

After yielding, the $\phi_{oil}=0.466$ sample flows with severe
shear thinning from $\sigma_Y$ to $\sigma\simeq0.3\,$Pa (Region 1,
Fig.~\ref{Fig2_FlowRegions} b). Depending on the region of fitting,
the flow curve may fit a power law,
\begin{equation}
{{\sigma} {\sim} \dot{\gamma}^{n_l} ,}\label{power_l}
\end{equation}
where $n_l\simeq 0.56$ when the data are fit over the range $1 \times
10^{-1}< \dot\gamma < 1 \times 10^{0}$, but $n_l\simeq 0.26$ if the
data are fit over the range $1 \times 10^{-2}< \dot\gamma < 1 \times
10^{-1}$. Hence the data do not reliably support one particular power
law. The shear thinning behaviour in region 1 exhibits a small yet
reproducible discontinuous jump in the shear rate, similar to a stress
plateau, that separates the different power law fits. This could be
related to wall slip or some complicated yielded banded flow as
recently proposed by Picard and co-workers \cite{Picard_PhysRevE_05}.
This type of complex behaviour could explain the difficulty in
reaching a steady state. An exponent of $n_l=0.6$ was predicted by
Meyer \textit{et. al.}  \cite{meyer} and Colby
\cite{Colby_EuroPhysLet_01} for defected lamellar phases by adapting
the Orowan equation for high temperature creep due to defects in
metals and alloys to describe plastic flow of defected lamellae.
Similar arguments were recently used in \cite{Ramos_PRL_2004} in a
phase of defected hexagonal micelles.

We will argue below that the stress step, or cliff, near
$\dot{\gamma}_c$ corresponds to a coexistence of lamellae and
onions. The putative onion-lamellae coexistence region spans
stresses $\sigma_l\simeq0.3\,$Pa to $\sigma_O\simeq1\,$Pa at a
strain rate $\dot{\gamma}_c\simeq1\,\textrm{s}^{-1}$ for
$\phi_{oil}=0.466$, followed by the homogeneous onion phase (region
2) at higher stresses. The onion phase shear thins according to
\begin{equation}
{{\sigma} {\sim} \dot{\gamma}^{n_o} ,}\label{power_o}
\end{equation}
where $n_o\simeq0.34$. This result agrees with studies on similar
systems
\cite{Courbin_PhysRevLett_04,Bergenholtz_Langmuir_1996,meyer,RND93}.
In the dilute sample the onion microstructure is apparently
destroyed at high stresses, and is replaced by a structure (region
3) whose behaviour is closer to Newtonian, ${\sigma} \sim
\dot{\gamma}$ \cite{Diat_JPhys_II_93_2}. The flow curve in the
transition between regions 2 and 3 is consistent with gradient shear
banding, but we have not studied this in detail. A stress plateau
would only be seen under controlled shear rate conditions . Another
possibility is vorticity banding in which the high shear rate phase
has a lower stress; in this case a negative apparent flow curve
could be seen under controlled strain rate conditions \cite{Olms99}.
The high shear lamellar phase is only measurable by the SR500
rheometer for the most dilute lamellar phase ($\phi_{oil}=0.644$).
For more concentrated systems the transition occurs at a higher
stress \cite{Diat_JPhys_II_93_2} and the sample was expelled from
the cone and plate geometry before this region was reached. We do
not know whether this is an inertial, surface, or normal stress
instability.
\begin{figure}[!ht]
\begin{center}
\includegraphics[width=3.5truein]{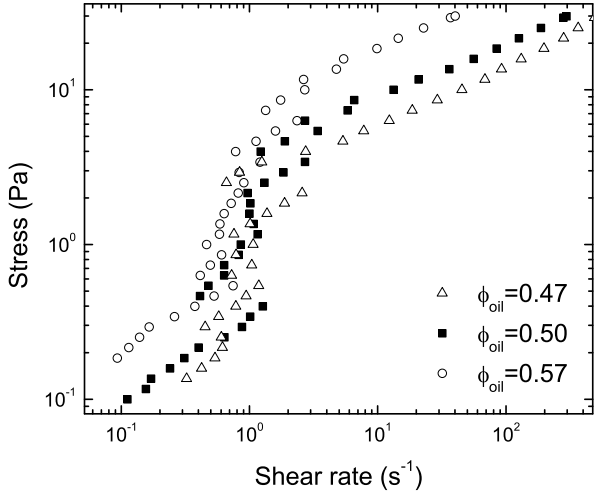}
\caption{Stress sweep flow curves for a range of oil
  concentrations. All the sweeps were performed using sweep criteria
  $\tau_{max}=2000\,\textrm{s}$ and $P=1\%$. }
\label{Fig3_SweepConc}
\end{center}
\end{figure}

The stress $\sigma_l$ at which the lamellae to onion transition
occurs decreases weakly with increasing oil fraction, or
equivalently decreasing layer spacing (Fig. \ref{Fig3_SweepConc}).
Hence, more dilute lamellar phases appear to require a lower stress
to induce the onion phase.

\begin{figure}[!ht]
\begin{center}
\includegraphics[width=3.5truein]{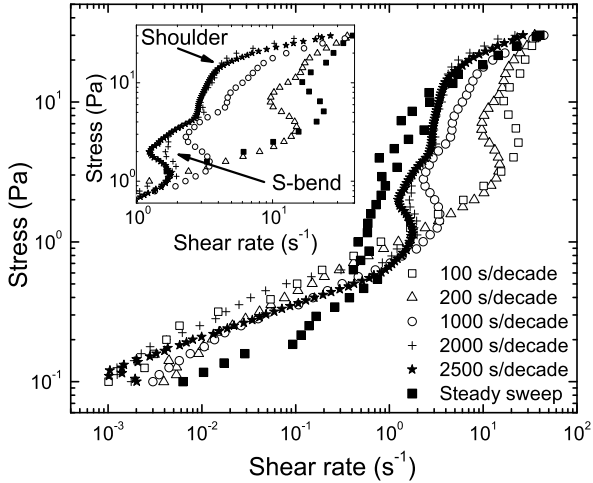}
\caption{Stress sweeps for $\phi_{oil}=0.47$. The time per stress
  increment is varied from $\tau_{max}=10-100$\,s and the number of
  stress increments per decade is varied from 10 to 50 points per
  decade, and the total time per decade is quoted on the plot.  Inset:
  enlargment near the \textsf{S} bend. As a comparison, data is also
  shown for a stress sweep with steady state conditions
  $\tau_{max}=2000\,\textrm{s}$ and $P=1\%$
  ($\scriptstyle\blacksquare$).}
\label{Fig4_TimeDep_pm47}
\end{center}
\end{figure}

The measured rheology between region 1 (lamellae) and region 2
(onions) depends upon the experimental protocol.
Figure~\ref{Fig4_TimeDep_pm47} shows stress sweeps at increasing
sweep times from $\tau_{s/d}=100\,$s to $\tau_{s/d}=2500\,$s where
$\tau_{s/d}$ is the sweep time calculated from the number of seconds
per decade of imposed stresses. The shear rate at which the stress
cliff occurs decreases as $\tau_{s/d}$ is increased. The
non-monotonic `\textsf{S}' bend present in all the non-equilibrium
flow curves. The \textsf{S} bend appears to shift to lower stresses
for slower sweeps, which implies that the lamellar phase can lose
stability quite slowly, and is broadly similar to nucleated
behaviour.  Stress sweeps with sweep times $\tau_{d/s}>200\,$s
exhibit two features in the shear rate cliff.  The first, already
noted, is the nonmonotonic \textsf{S} bend and occurs at the
termination of region 1. The second is the vertical `shoulder' which
occurs before the beginning of region 2. The shoulder is most
prominent for the slowest sweeps.  The slowest sweeps
($\tau_{s/d}=2000\,$s and $\tau_{s/d}=2500\,$s) almost superpose
during the step region, with the \textsf{S} bend and the shoulders
occur at similar shear rates.

The slope of the low stress lamellar phase branch changes with the
sweep time. The branches are fit with the power law $\sigma \sim
\dot\gamma^{n_l}$ where the exponent $n_l$ ranges from $0.235 < n_l
< 0.342$ depending on the sweep time (Table~\ref{timedep}). The
slowest stress sweeps shear thin with the smallest exponent,
$n_l\sim0.235$, which could indicate that fewer defects have been
produced or that more defects have been removed. The transformation from
lamellae to onions is thus very slow. A couple of possible
mechanisms could be: (1) a Helfrich-Hurault like undulatory
instabilty that grows slowly and takes a long time to convert to
onions; or (2) a slow nucleation process. The flow curves measured
for the slow stress sweeps reach the onion flow branch at lower
shoulder stresses ($\sigma_{sh}$) as shown in the table and figure.
Perhaps this is due to the slow conversion to onions, such that for
fast sweeps the conversion occurs at larger $\sigma_{sh}$.

We can also interpret the \textsf{S} bend described above as a
feature associated with the formation of onions. The backwards
\textsf{S} is consistent with progressive stress inducing more
onions hence decreasing the shear rate. The forwards \textsf{S} at
higher stresses implies that any contribution to the measured shear
rate from conversion of lamellae to onions (which would lead to
decreasing shear rate) is overwhelmed by the tendency of the onions
to act normally and flow faster. Since there are many more onions at
higher stresses their effect is far greater.

\begin{table}
\begin{center}
\caption{Power law exponents for the lamellar flow
  branch from figure \protect{\ref{Fig4_TimeDep_pm47}}, together with
  the sweep time per decade ($\tau_{s/d}$) and the stress
  $\sigma_{\textrm{sh}}$ and shear rate
  $\dot{\gamma}_{\textrm{sh}}$ at the top of the shoulder in the
  step region.}
\label{timedep}
\begin{tabular}{llll}
\hline\noalign{\smallskip}
  $\tau_{s/d}$ & $\sigma_{\textrm{sh}}$ (Pa) &
  $\dot\gamma_{\textrm{sh}}$ (s$^{-1}$)& $n_l$ \\
\noalign{\smallskip}\hline\noalign{\smallskip}
  100 & 30.0 & 40.3 & 0.342 \\
  200 & 23.9 & 16.9 & 0.338 \\
  1000 & 19.1 & 10.7 & 0.283 \\
  2000 & 16.3 & 4.2 & 0.246 \\
  2500 & 16.3 & 4.2 & 0.235 \\
\noalign{\smallskip}\hline
\end{tabular}
\end{center}
\end{table}

\section{Evidence for vorticity shear banding}\label{sec:4}
\subsection{Flow segments and
  discontinuous shear rate jumps}

\begin{figure}[!ht]
\begin{center}
\includegraphics[width=3.5truein]{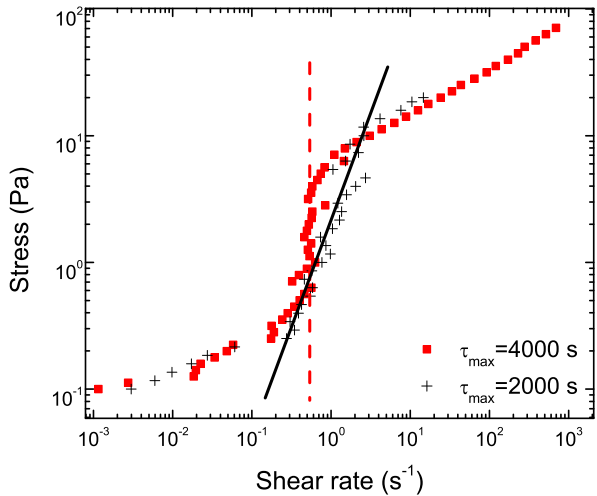}
\caption{Stress sweeps for $\phi_{oil}=0.50$. Steady state
conditions
  were different for the two sweeps; when
  $\tau_{max}=2000\,\textrm{s}$, $P=1\%$ and when $\tau_{max}=4000\,\textrm{s}$,
  $P=0.1\%$.}
\label{pm50_timedep}
\end{center}
\end{figure}

\begin{figure}[!ht]
\begin{center}
\includegraphics[width=3.5truein]{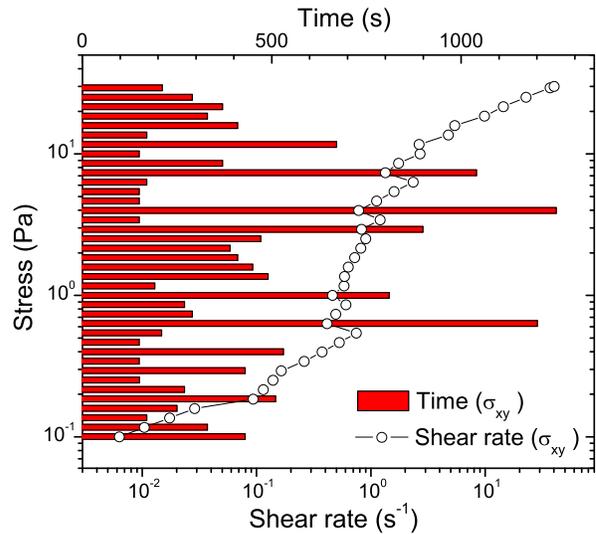}
\caption{Stress sweep for $\phi_{oil}=0.466$ using the steady state
  criteria $\tau_{max}=2000$ s and $P=1\%$.  The horizontal bars
  denote the time the material required to reach a steady state within
  the criterion $P=1\%$.
\label{fig:Fig5_TimeGap_p46_091203}}
\end{center}
\end{figure}

\begin{figure}[!ht]
\begin{center}
\includegraphics[width=3.5truein]{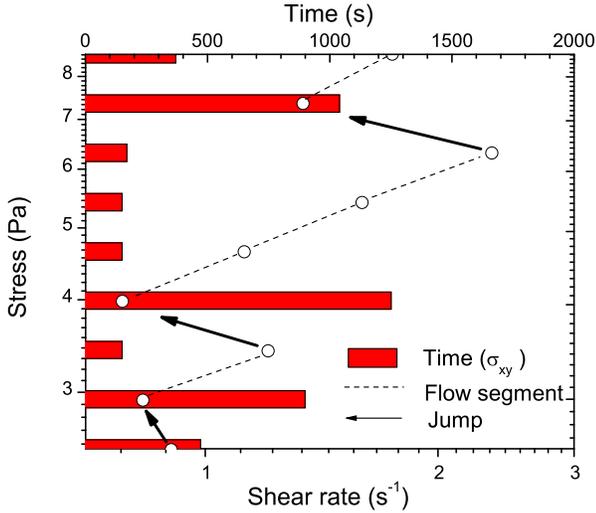}
\caption{Enlargement of the top of the step region in figure
  \ref{fig:Fig5_TimeGap_p46_091203}. The flow segments are shown as
  dotted lines and the jumps are shown with arrows.}
  \label{Fig6_TimeGap_p46_091203}
\end{center}
\end{figure}
\begin{figure*}[!htb]
  \begin{center}
    \includegraphics[scale=0.4]{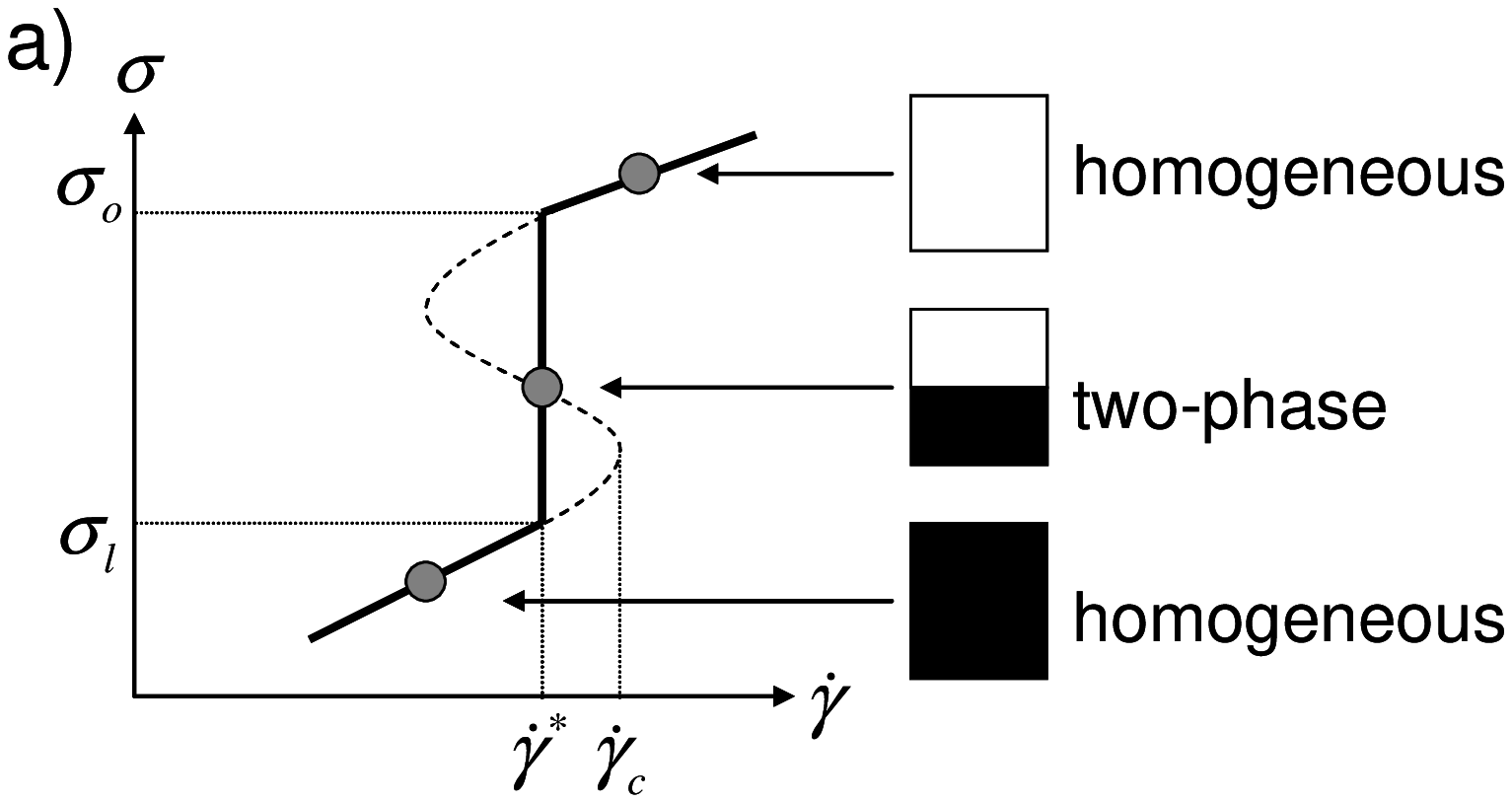}
    \includegraphics[scale=0.4]{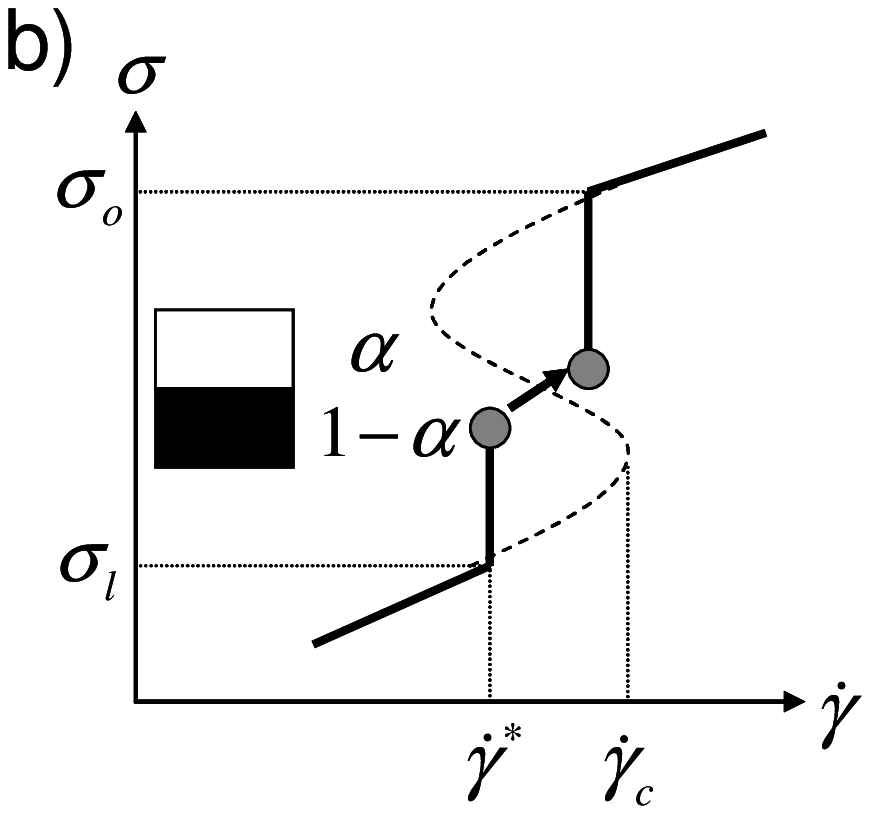}
    \includegraphics[scale=0.4]{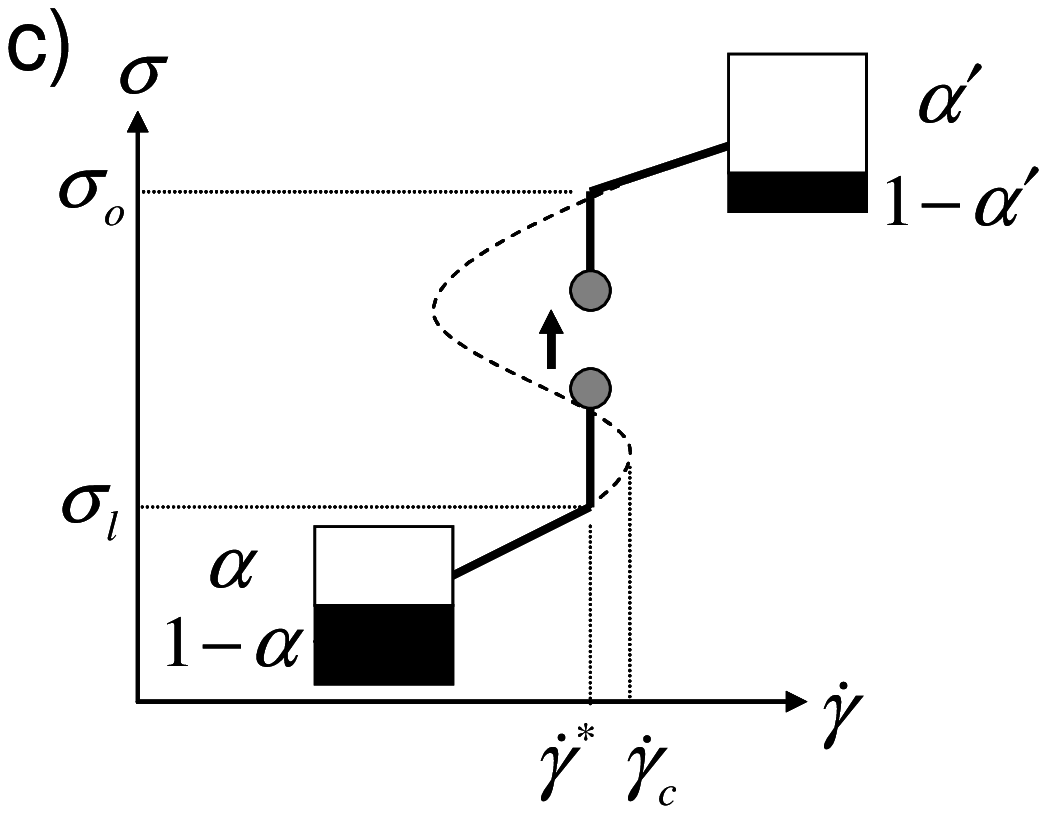}
    \caption{a): Flow curve showing three possible states; two
      homogeneous and one vorticity banded, in an ideal situation in
      which the two states coexist at a ``selected'' shear rate
      $\dot{\gamma}^{\ast}$.  Two possible scenarios are shown for the
      evolution of a vorticity-banded configuration upon increasing
      the stress.  In b) the fraction of material $\alpha$ in the onion
      branch stays the same, and the two phases evolve
      along their respective flow branches, hence increasing the shear
      rate $\dot{\gamma}$ above $\dot{\gamma}^{\ast}$. In c) $\alpha$
      increases to $\alpha'$ after which the system returns to the
      selected strain rate $\dot{\gamma}^{\ast}$.}
    \label{fig:Fig10_cartoon}
  \end{center}
\end{figure*}

We have shown above that the measured flow curve is very senstive to
the sweep time in the region of the lamellar to onion transition, in
the step region. In this section we focus in more detail on the
rheology in this region.  Stress sweeps were performed with 15 stress
increments per decade and a maximum stress step time of
$\tau_{max}=2000\,\textrm{s}$ giving a maximum sweep time of
$\tau_{d/s}=30000\,\textrm{s}$, which is significantly longer than the
sweep times applied above (Fig.~\ref{Fig4_TimeDep_pm47}). The steady
state condition was set to $P=1\%$.

Figure \ref{pm50_timedep} presents flow curves with sweep criteria
$\tau_{max}=2000\,\textrm{s}$ and $\tau_{max}=4000\,\textrm{s}$. It
is clear that slower sweeps (larger $\tau_{max}$) produces a more
vertical step region.  However, increasing $\tau_{max}$ does not
significantly alter the flow curves in region 1 and region 2; this
contrasts what we reported for fixed time stress sweeps (see
Fig.~\ref{Fig4_TimeDep_pm47}) where altering the sweep time
($\tau_{d/s}$) affected both the slope of the flow curve in region 1
and the step region. Furthermore, the critical shear rate
$\dot\gamma^\ast$ is similar for both sweeps regardless of the
duration of $\tau_{max}$ ($\dot\gamma^\ast=0.31\,\textrm{s}^{-1}$
for the faster sweep and $\dot\gamma^\ast=0.51\,\textrm{s}^{-1}$ for
the slower).

Figure \ref{pm50_timedep} indicates jagged behaviour during the step
region, during which the measured shear rate oscillates between
$\dot\gamma^\ast$ and a larger shear rate. The fast stress sweep is
more jagged than the slow sweep. On closer inspection of the step
region one notices a series of intermediate flow segments with slopes
slightly steeper than the lamellar constitutive curve, separated by
discontinuous jumps to lower shear rates as the stress is increased.

Figures \ref{fig:Fig5_TimeGap_p46_091203} and
\ref{Fig6_TimeGap_p46_091203} show the measured flow curves for a
stress sweep of a sample with $\phi_{oil}=0.466$. The jagged
behaviour described earlier is apparent in the step region. Also
plotted is the steady state time $\tau_{ss}$, which is the time
neccessary to reach the steady state condition $P=1\%$. Along the
lamellar and onion flow branches $\tau_{ss}$ is always a few hundred
seconds, and remains fairly constant for subsequent stress
increments.  However, during the step region $\tau_{ss}$ varies
widely, $150\,\textrm{s} < \tau_{ss} < 1200\,\textrm{s}$.  There is
a strong correlation between $\tau_{ss}$ and the jagged shape of the
step region.  Long steady state times occur for jumps, \textit{i.e.}
stress steps that lead to a decrease in shear rate; while the
shorter steady state times correspond to moving along the flow
segments and increasing the strain rate.  This correlation between
$\tau_{ss}$ and jumps and flow segments occurs for stress sweeps of
all the concentrations studied and is reproducible.

The step region itself is broadly consistent with vorticity banding
between the queiscent lamellar phase and thicker flow-induced onion
phase, and we will argue below that the jagged behaviour is due to
very slow and complex transformation kinetics between lamellar and
onion states.
\subsection{Model for stress evolution during vorticity banding}
In a vorticity banded state the total imposed stress $\bar{\sigma}$
must equal the sum of the individual stresses in the shear bands,
\begin{equation}
    \bar{\sigma}(\dot{\gamma}) = \alpha{\sigma}_{O}(\dot{\gamma}) +
    (1-\alpha){\sigma}_{L}(\dot{\gamma}), \label{constitutive}
\end{equation}
where $\alpha$ is the fraction of material in the onion state
(converted from lamellae) and $\sigma_{O}(\dot{\gamma})$ and
$\sigma_{L}(\dot{\gamma})$ are the constitutive relations of the
onion and lamellae shear band respectively. In true steady state the
shear rate $\dot{\gamma}^{\ast}$ at which coexistence between
lamellae and onions can occur may be uniquely selected
\cite{olmsted99c,olmgov01}, by analogy with the shear stress that is
apparently selected in gradient banding \cite{olmsted92,LOB00}. For
materials whose coexisting states have the same concentration the
selected shear rate is predicted to be independent of concentration,
resulting in a stress ``cliff''; while materials in which the
coexisting state have different concentrations may have a selected
shear rate $\dot{\gamma}^{\ast}$ that depends on stress, or
equivalently on $\alpha$ \cite{olmsted99c,Olms99}.

\begin{figure*}[!ht]
\begin{center}
    \includegraphics[scale=0.55]{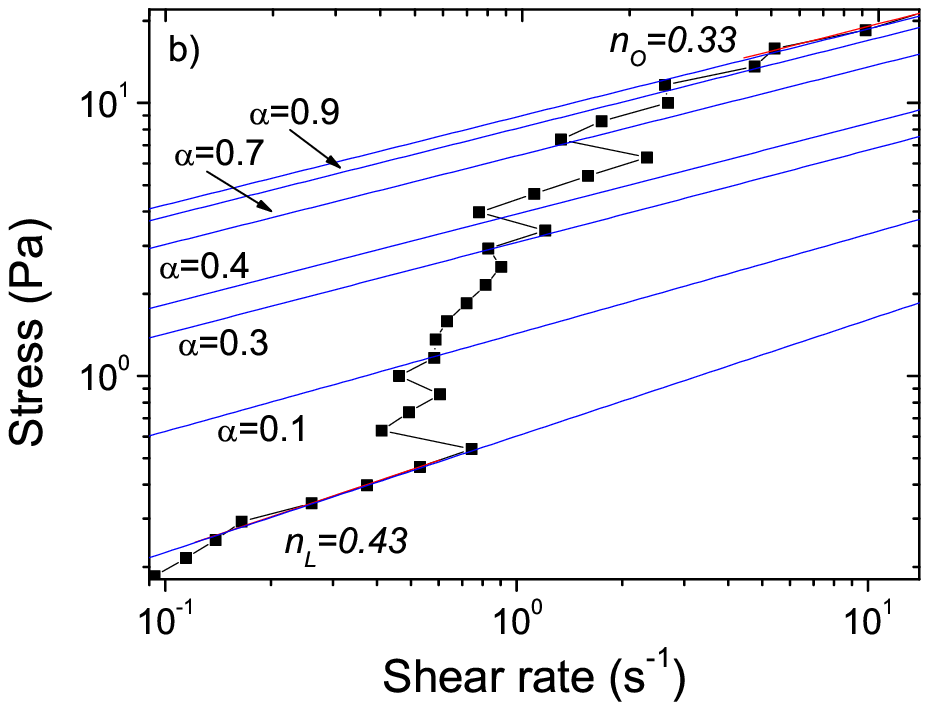}
    \includegraphics[scale=0.55]{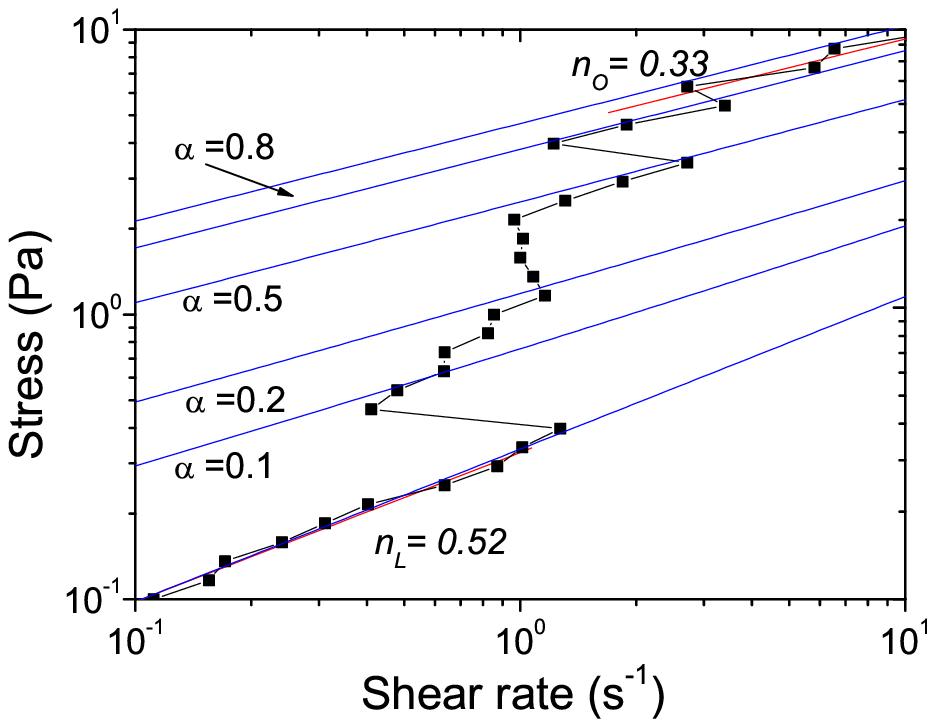}
    \includegraphics[scale=0.55]{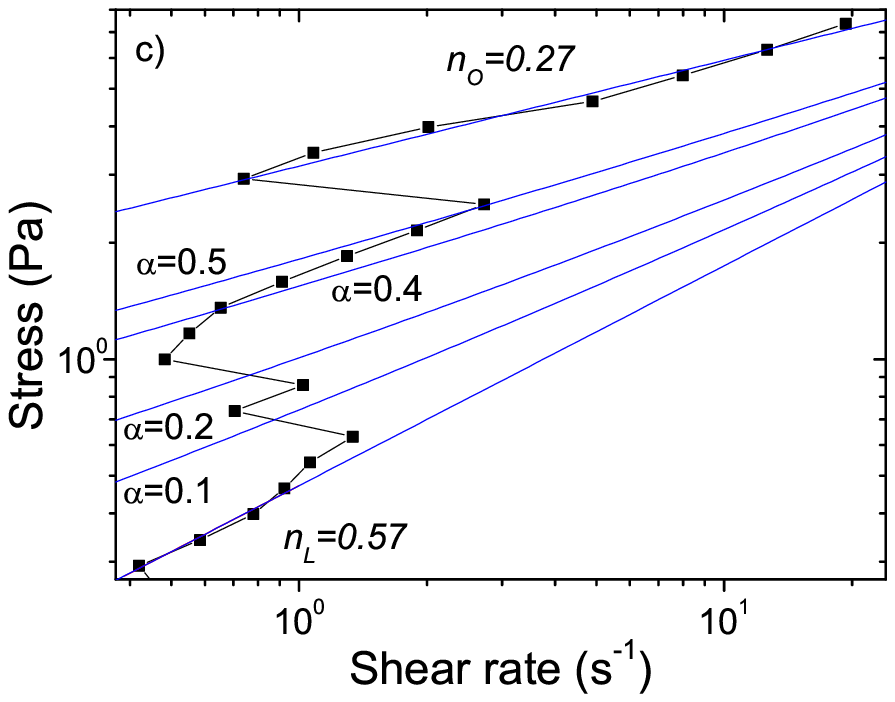}
\caption{Stress sweeps using steady state criteria
  $\tau_{max}=2000\,\textrm{s}$ and $P=1\%$. Power laws are plotted
  with different $\alpha$ values according to Eq.~\ref{constitutive}:
  a) $\phi_{oil}=0.466$, b) $\phi_{oil}=0.50$, c) $\phi_{oil}=0.569$.} \label{Fig8_fixed}
\end{center}
\end{figure*}

\begin{figure*}[!ht]
\begin{center}
\includegraphics[scale=0.8]{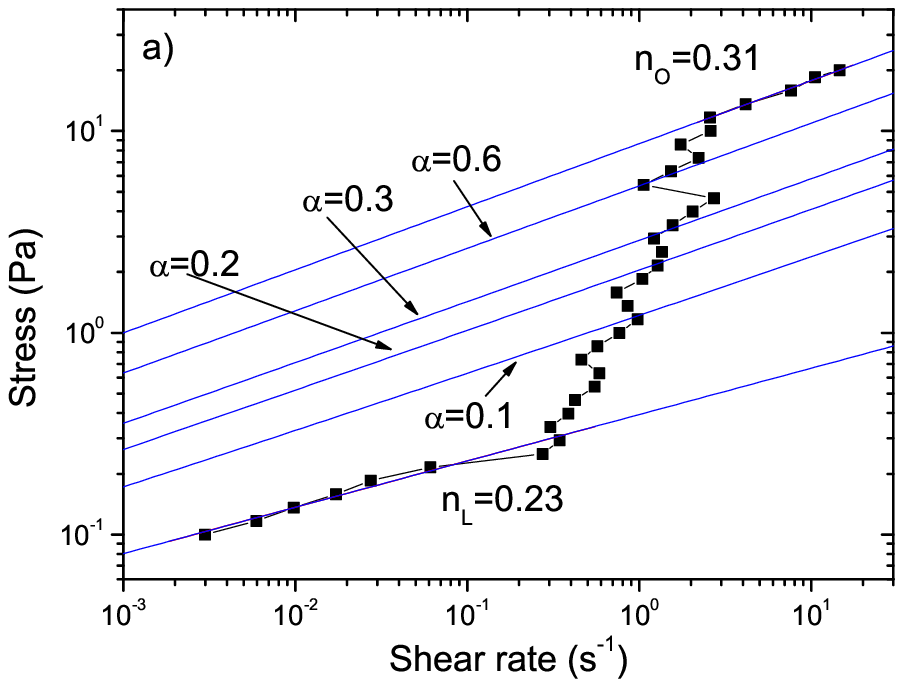}
\includegraphics[scale=0.8]{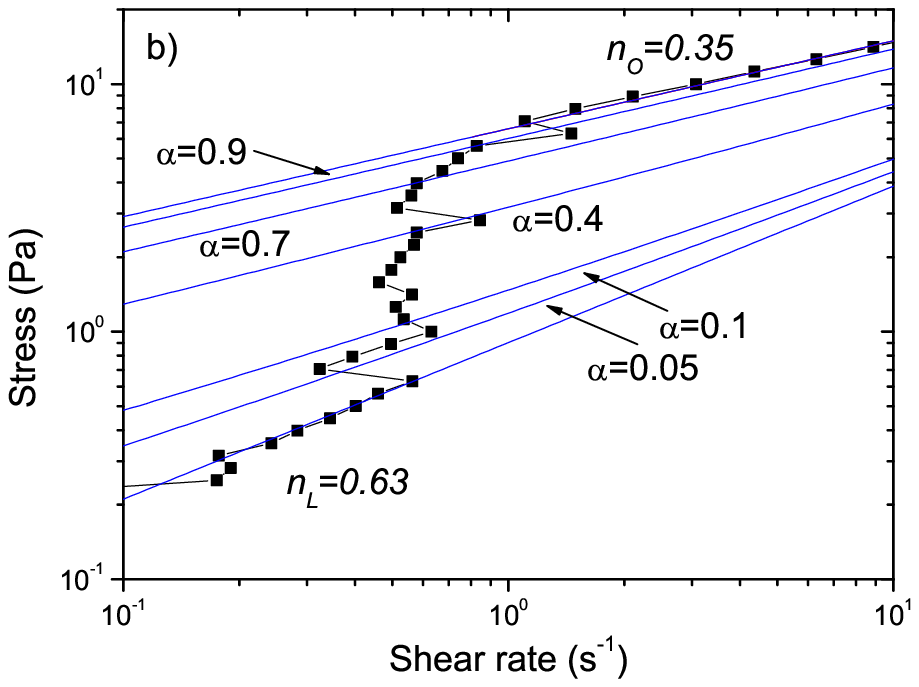}
\caption{Stress sweeps using different  steady state
  criteria;a) $\tau_{max}=2000\,\textrm{s}$ and $P=1.0\%$, b)
  $\tau_{max}=4000\,\textrm{s}$and $P=1.0\%$. Both sweeps were
  carried out on lamellar phases with oil concentration
  $\phi_{oil}=0.50$.
\label{fig:fixed}}
\end{center}
\end{figure*}

For the sake of argument, we first suppose that lamellar-onion
coexistence has an ideal constant selected shear rate
$\dot{\gamma}^{\ast}$ independent of stress.  Consider a
vorticity-banded state at a given stress $\sigma$ and shear rate
$\dot{\gamma}^{\ast}$, with some fraction $\alpha$ of onion
material. Upon increasing the stress there are two possibilities:
\begin{itemize}
\item[(i)] $\alpha$ can remain fixed, with the onion and lamellar
  phases increasing their shear rates to accommodate the increased
  stress, and hence following their respective constitutive curves.
  Successive stresses would lead to a flow branch intermediate between
  the lamellar and onion branches, given by Eq.~(\ref{constitutive}),
  as shown in Fig.~\ref{fig:Fig10_cartoon}b.

\item[(ii)] $\alpha$ can increase as lamellae convert to onions and
  the shear rate eventually returns to the selected value
  $\dot{\gamma}^{\ast}$. This could occur if the lamellar phase no
  longer ``absorbs'' the increased strain rate demanded by the stress,
  but responds either by an instability of by nucleating new onion
  material (Fig.~\ref{fig:Fig10_cartoon}c).
\end{itemize}
Scenario (i) is consistent with the evolution along flow segments
that we find, while scenario (ii) is consistent with the jumps
between flow segments.  To test this we have extracted the rheology
of the lamellar ($\sigma_l(\dot{\gamma})$) and onion
($\sigma_O(\dot{\gamma})$) branches, and shown the expected rheology
of a heterogenous mixture that evolves at constant onion fraction
$\alpha$, according to Eq.~(\ref{constitutive}), in
Fig.~\ref{Fig8_fixed} for different onion fractions $\alpha$. It is
clear that some of the flow segments follow the fixed $\alpha$ flow
branches quite well. Fig.~\ref{fig:fixed} shows sweeps with
different steady state times $\tau_{ss}$; it is apparent that the
two experiments yield similar behavior, but the onset transition is
at a lower strain rate for the slower experiment.

This simple picture would, of course, be complicated by very slow
transformation kinetics between states. Moreover, it appears that
the coexistence strain rate, if there is one, is not constant but
slowly increases with stress. This might be expected for a solution,
in which the coexistence conditions can be expected to depend on
concentration, which leads to a stress-dependent selected strain
rate because of a biphasic window in the equivalent non-equilibrium
strain rate-concentration phase diagram \cite{Olms99,OlmsLu97}. We
stress the fact that no such phase diagram has been calculated for a
lyotropic lamellar system, and that these data are merely consistent
with such a phase diagram.

This behaviour of the L-O transition should be contrasted with the
well-known kinetic behaviour seen in gra\-dient banding fluids such as
wormlike micelles \cite{BerrPort99,Berr97,grand97,lerouge2}.
Wormlike micelles can phase separate into two bands along a stress
plateau for a range of shear rates (Fig.~\ref{fig:bandcartoon}a).
Upon increasing the shear rate while on the plateau the systems
invariably return to the stress plateau, corresponding to an increase
in the fraction $\alpha$ of the high shear rate phase \cite{lerouge2}.
By contrast , the SDS mixture we study appears to get ``stuck'' in the
low stress lamellar phase much more easily than the micellar phase
gets stuck in the low shear rate phase. There are several possible
contributing factors for this:
\begin{enumerate}
\item In a vorticity-banded state the interface lies in the shear
  plane, while in a gradient banded state the shear flow occurs
  across, rather than within, the interface. The lack of a direct
  shear force across the interface could lead to slower interface
  motion.
\item The lamellar and onion phases are not smoothly related by a
  continuous order parameter; hence one may expect that
  nucleation-like behavior, should it occur under flow, would be very
  slow. One candidate for onion formation is a critical shear rate
  for an undulation instability; it is possible that this instability
  can be preempted by finite amplitude fluctuations which can nucleate
  onions below the critical shear rate for the instability, but that
  these fluctuations are very slow. In the wormlike micellar systems
  such a nucleation barrier, if there is one, is apparently relatively
  easy to overcome \cite{BerrPort99,Berr97,grand97,lerouge2}.
\item Vorticity banding would present an interface that has no
  preferred position with reference to the flow geometry. By contrast,
  gradient banding in cylindrical Couette or cone and plate geometries
  leads to an interface that lies in a stress gradient set by the
  curvature of the flow. This stress gradient effectively drives the
  interface to move until it is at the position of the selected stress
  $\sigma^{\ast}$ \cite{olmsted99a,lerouge2}. There is no such driving
  force for vorticity banding, which again leads to slow kinetics.
\end{enumerate}

In the steady state tests the times between data points provide
further evidence for this interpretation
(Fig.~\ref{fig:Fig5_TimeGap_p46_091203} shows the time to reach
steady state $\tau_{ss}$ as bars). During flow segments each stress
step requires only a few hundred seconds at most before the steady
state criteria are satisfied. The discontinuous shear rate ``jump''
to a lower shear rate takes a long time to reach a steady state
within the set criteria. Flow segments are thought to be associated
with no or very small changes in microstructure, and shear rate
jumps related to events where lamellae are converted to onions.
Hence, the time to reach a quasi-steady state at fixed $\alpha$
should be governed by the processes associated with moving along the
homogeneous flow branches of the lamellar and onion phases. In the
lamellar phase this involves changing the steady state defect
density \cite{meyer}, while in the onion phase the onion size should
decrease \cite{Diat_JPhys_II_93_2}. The very large steady state time
indicates a rapid initial change in conditions upon the first
thickening event, followed by a slow relaxation to steady state.
This would include, in the case of onion formation associated with
an increased $\alpha$, the formation of an entire new band of onion
phases around the entire circumference of the rheometer, as well as
a return of all onion sizes to the larger size expected at the lower
shear rate.

To summarise: we suggest that the flow segments correspond to an
increasing stress in which the fraction of onion material remains
fixed, separated by discrete jump decreases in shear rate that
correspond to the formation of new bands of onions.  We attribute
the slow and irregular /non-reproducible behaviour to the
nucleation-like behaviour of onion formation for small increases in
stress.

\section{Conclusions}
In this work we have studied the rheology of a lyotropic lamellar
system undergoing a transition between and lamellar and
multi-lamellar, or ``onion'', phases.  The transition is
discontinuous, and the flow curves $\sigma\simeq\dot\gamma^n$ follow
distinct forms in the different phases: in the lamellar regime
$\sigma\sim\dot\gamma^{0.6}$ (depending on the region of fitting),
and in the onion regime $\sigma\sim\dot\gamma^{0.3}$. The latter is
consistent with previous measurements \cite{Roux_EuroPhysLett_93},
while the former is consistent with theory and experiment
\cite{Meyer_JPhysEur_2001}. We have also identified a very low
strain rate signature consistent with yield due to a defect network,
from which we estimate defect spacings consistent with previous work
on lyotropic lamellar phases \cite{meyer}.

The discontinuity in the measured shear stress at a given shear rate
rate is a ``cliff'', analogous to the stress plateau found in the
better understood wormlike micelle system. In shear thinning wormlike
micelles the stress plateau indicates macroscopic phase separation
into bands of material flowing with different local shear rates; while
the cliff in the lamellar/onion case is consistent with bands lamellae
and onions stacked in bands along the vorticity axis, which have the
same shear rate but different shear stresses. Using a simple model for
the superposition of the stresses in the two phases at coexistence, we
are able to understand the noisy behaviour along the plateau as due to
nucleation (or instability and subsequent slow growth) of onions out
of the lamellar phase. Because nothing is expected to break the
symmetry along the vorticity axis, we do not expect two large
macroscopic bands (of the two phases), but rather a much finer
dispersion of bands.  Although other groups have reported vorticity
banding in colloidal suspensions \cite{ChenZAHSBG92,ChenCAZ94} and
wormlike micelles \cite{WFF98,callaghan}, this is the first explicit
attempt to understand the rheology of such a phase separated state.

In another previous study of a lyotropic lamellar system, Bonn
\textit{et al.}  visualized vorticity banding in AOT, during transient
experiments \cite{Bonn+98}. The flow curves in that case exhibited
apparent shear thinning, and it is not apparent how the reported bands
related to the overall steady state of the materials. Nonetheless,
there are probably similar phenomena occuring, at least in part, in
the two systems.

Our results thus strongly suggest that the lamellar to onion
transition for the quaternary mixture of SDS, water, pentanol and
dodecane exhibits vorticity shear banding.  Direct observation of
the lamellar phase in the vorticity-flow plane during stress
controlled shear is necessary to confirm this hypothesis along with
further study on the microstructure of the shear bands and this work
is currently in progress. Initial birefringence studies indicate
vorticity shear bands of differing anisotropies exist in the step
region.

We thank the UK EPSRC for financial support, and R. Colby, S.
Manneville, J. Dhont and P. Lettinga for advice. We are especially
grateful to S. Lerouge for invaluable guidance during the early
 stages of this work.

\end{document}